\documentclass[12pt]{article}
\usepackage[cmtip,arrow]{xy}
\usepackage{pb-diagram,pb-xy}
\usepackage{amsmath}
\usepackage[psamsfonts]{amssymb}
\usepackage{cmmib57}
\usepackage{amsmath,amssymb,amscd}
\usepackage{tikz-cd}
\newcommand{\bPf}{\par\vspace*{-4pt}\indent{\sc Proof.}\enskip}
\newcommand{\ePf}{\medskip}
\def\QED{\hskip0.1em\hfill\null\ \null\nobreak\hfill\kern3pt\vbox{\hrule\hbox
   {\vrule\kern1pt\vbox{\kern1.7pt\hbox{$\scriptscriptstyle{QED}$}
    \kern0.2pt}\kern1pt\vrule}\hrule}}

\def\END{\hskip0.1em\hfill\null\ \null\nobreak\hfill\kern3pt\vbox{\hrule\hbox
   {\vrule\kern1pt\vbox{\kern1.7pt\hbox{$\,\,\,\vspace{5pt}$}
    \kern0.2pt}\kern1pt\vrule}\hrule}}
\newtheorem{theorem}{Theorem}[section]
\newtheorem{lemma}[theorem]{Lemma}
\newtheorem{corollary}[theorem]{Corollary}
\newtheorem{proposition}[theorem]{Proposition}
\newtheorem{remark}[theorem]{Remark}
\newtheorem{definition}[theorem]{Definition}
\newtheorem{example}[theorem]{Example}

\newcommand{\bCd}{\beq \begin{CD}}
\newcommand{\eCd}{\end{CD}\eEq}
\newcommand{\bcd}{\beq \begin{CD}}
\newcommand{\ecd}{\end{CD}\eeq}
\newcommand{\ben}{\begin{enumerate}}
\newcommand{\een}{\end{enumerate}}
\newcommand{\bEq}{\begin{eqnarray}}
\newcommand{\eEq}{\end{eqnarray}}
\newcommand{\beq}{\begin{eqnarray*}}
\newcommand{\eeq}{\end{eqnarray*}}
\newcommand{\bDf}{\begin{definition}\em}
\newcommand{\eDf}{\end{definition}}
\newcommand{\bLm}{\begin{lemma}}
\newcommand{\eLm}{\end{lemma}}
\newcommand{\bPr}{\begin{proposition}}
\newcommand{\ePr}{\end{proposition}}
\newcommand{\bTh}{\begin{theorem}}
\newcommand{\eTh}{\end{theorem}}
\newcommand{\bCr}{\begin{corollary}}
\newcommand{\eCr}{\end{corollary}}
\newcommand{\bRm}{\begin{remark}\em}
\newcommand{\eRm}{\end{remark}}
\newcommand{\bEx}{\begin{example}\em}
\newcommand{\eEx}{\end{example}}

\newcommand{\ie}{{\em i.e$.$} }
\newcommand{\eg}{{\em e.g$.$} }

\newcommand{\mto}{\mapsto}

\newcommand{\cI}{\mathcal{I}}
\newcommand{\cJ}{\mathcal{J}}

\newcommand{\bY}{\boldsymbol{Y}}

\title{\large{ {\bf
The Jacobi morphism and the Hessian in higher order field theory;
 with applications to a Yang--Mills theory on a Minkowskian background}}}
\author{{\normalsize Luca Accornero
}
\\
{\footnotesize Department of Mathematics, University of Utrecht}
\\
{\footnotesize 3508 TA Utrecht, The Netherlands}
\\  
{\footnotesize e--mail: 
{\sc l.accornero@uu.nl
}}
\\
{\normalsize Marcella Palese}
\\ {\footnotesize Department of Mathematics,
University of Torino}
\\
{\footnotesize via C. Alberto 10, 10123 Torino, Italy} 
\\  {\footnotesize e--mail: 
{\sc marcella.palese@unito.it}}}
\date{}
\begin{document}
\maketitle

\begin{abstract}
We characterize the second variation of an higher order Lagrangian by a Jacobi morphism and by 
currents strictly related to the geometric structure of the variational problem. We discuss the relation between
the Jacobi morphism and the Hessian at an arbitrary order. Furthermore, we prove that a pair of Jacobi fields always generates a (weakly) conserved current.  An explicit example is provided for 
a Yang--Mills theory on a Minkowskian background.
\end{abstract}



\section{Introduction}

We focus on the study of second 
variations of Lagrangians on finite order prolongations of fibered manifolds \cite{Eh51b-52,Sau89}, a natural, geometic framework for the calculus of variation \cite{AlAz78,An,Ec81,GoSt73,KMS93,Kru73,Kup80,Sau89,Tra67}.
Within this formulation the higher variations can be interpreted as variations of suitable `deformed' Lagrangians.
This fact appears to be of interest in theoretical physics, in particular concerning issues on variations of currents \cite{FrPaWi13}; see for example some results dealing with applications of the second variation in the theory of gravitation \cite{FeFrRa03}. 

An important intrinsic feature of prolongation spaces is the contact structure and in particular, the so called finite order contact ideal \cite{Kru90,Kru15}.
Such a structure allows one to introduce an geometric integration by parts procedure on differential forms; this procedure enables to treat variation formulae intrinsically. In what follows we refer to Krbek and Musilov\'a 's use of the so called interior Euler operator  \cite{Kr02,KrMu03,KrMu05}; for different approaches see \eg \cite{Kol83,FF03}.

Indeed, we recall that by the contact  structure, a differential-geometric frame for the calculus of variations  - the finite order (exact) variational sequence \cite{Kru90} - is obtained as the quotient, by a suitably defined contact subsequence, of the de Rham sequence of sheaves of forms and, in fact, the use of the interior Euler operator has been introduced in order to obtain a representation by differential forms of the variational sequence (called the Takens representation); see \eg \cite{FFP01,KrMu05,KrUrVo13,PaRoWiMu16,VoUr14} and, for applications in topological obstructions in Lagrangian field theories, \cite{PaWi17}. 

In this paper we deal with the interior Euler operator as a tool to discuss variation formulae of higher order and their associated conservation laws.
The second order case is of special interest for our purposes.
Indeed, in Section \ref{Jacobisection} the second variation is analyzed in detail: we define the Jacobi morphism in terms of the interior Euler operator, recovering several properties discussed in \cite{GoSt73} for first order field theories, \ie self adjointness along 
critical 
solutions and the relation with the Hessian of the action functional; here such properties are extended to higher order field theories.
Moreover, we discuss strong and weak conservation laws associated with Jacobi fields; in particular, as a main result, we show that to any pair of Jacobi fields corresponds a weakly conserved current. 

The explicit example of the Jacobi equation for a Yang-Mills theory on a Minkowskian background is worked out in Section \ref{Yang--Mills}. Our result can be compared with the classical definition of a Jacobi operator for Yang--Mills Lagrangians, see \eg \cite{AtBo83,Bou87}, adapted to our specific case.
Furthermore, we explicate the relation between Jacobi fields, symmetries of 
variations and conserved currents. 
In particular, the current associated with two Jacobi fields is obtained for a Yang-Mills theory on a Minkowskian background.

\subsection{Jet prolongations and the finite order contact structure}\label{section2}

We briefly recall the modern geometric approach to calculus of variations on finite order prolongations of fibered manifolds. 
We denote by $X$ a differentiable manifold of dimension $n$ and by $Y$ a differentiable manifold of dimension $m+n$; we assume that it exists a fibered manifold structure $(Y,\pi,X)$ in which $X$ is the base space, $Y$ is the total space and $\pi$ is the projection. Only local fibered coordinates, \ie adapted to the fibration, $(x^i,y^\sigma)$, with $i=1\dots n$ and $\sigma=1\dots m$, will be used. In what follows we shall denote $ds=dx^1\wedge\ldots\wedge dx^n$ the local expression of a volume element on $X$; furthermore, we use the following notation
$
ds_i=\frac{\partial}{\partial x^i}\rfloor ds$,  $
ds_{ij}=\frac{\partial}{\partial x^i}\rfloor \frac{\partial}{\partial x^j}\rfloor ds $,  $\dots $, and so on.

An equivalence relation identifies  local sections of $\pi$ defined in a neighborhood of $x\in X$ such that they have the same values and derivatives, up to the order $k$, at $x$. The $k$-th jet space $J^kY$ is defined as the space of such equivalence classes which we denote by $j^k_x\phi$. We have, for $0\leq h<k$, a map $\pi_{k,h}: J^kY \to J^h\pi$ such that $\pi_{k,h}(j^k_x\phi)=j^h_x\phi$, where we set $J^0\pi=Y$ and $j^0_x\phi=\phi(x)$. It turns out that $J^kY$ is a manifold; moreover, the maps $\pi_k: J^kY \to X$ defined by $\pi_k(j^k_x\phi)=x$ turn out to be surjective submersions; therefore, the triples $(J^kY,\pi_k,X)$ are fibered manifolds.
It can be proved that, for every $k$ and every $0\leq h\leq k$, $(J^kY,\pi_{k,h},J^h\pi)$ are fibered manifolds as well and that $(J^kY,\pi_{k,k-1},J^{k-1}\pi)$ is always an affine bundle. 

For $0 \leq h\leq k$, by setting
$
y^\sigma_{j_1\dots j_h}(j^k_x\phi)=\left.\frac{\partial^{h}\phi^\sigma}{\partial x^{j_1}\dots \partial x^{j_h}}\right|_x $, we have a set of functions $(x^i, y^\sigma, y^\sigma_{j_1}\dots ,y^\sigma_{j_1\dots j_k})$ defined locally on $J^kY$; restricting to $j_t\leq j_q$ if $t\leq q$ they are a proper coordinates system.  

Given a section of $\pi$, denoted by $\sigma$, the $k$-prolongation $j^k\sigma$ is a section of $\pi_k$
defined by
$
y^\alpha_{j_1\dots j_h}(j^k\sigma)=\frac{\partial^{h}\phi^\alpha}{\partial x^{j_1}\dots \partial x^{j_h}}
$. Sections of $\pi_k$ that are not of this type are called non holonomic sections.

Finally, given a function $f$ defined on an open set $V$ of $J^kY$ and an index $1\leq i\leq n$ the $i$-th {\em total derivative} is a function defined on $\pi^{-1}_{k+1,k}(V)$ with expression
$
d_i f=\frac{\partial f}{\partial x^i}+\sum\limits^{k}_{t=1}\sum\limits_{j_1\leq \dots \leq j_t}\frac{\partial f}{\partial y^\sigma_{j_1\dots j_t}}y^\sigma_{j_1\dots j_t i}$.
We shall use the following convention on multi-indices:
\begin{itemize}
\item[-] a multi-index will be an ordered $s$-uple $I=(i_1, \dots i_s)$;
\item[-] the length of $I$ is given by the number $s$;
\item[-] an expression of the kind $Ij$ denotes the multi-index given by the $(s+1)$-uple $(i_1, \dots i_s, j )$.
\end{itemize}
Therefore, we have
$
\frac{\partial^{|I|}}{\partial x^I}=\frac{\partial^{s}}{\partial x^{i_1}\dots \partial x^{i_s}}$. If $1\leq i_1 \dots \leq i_s\leq n$ the system $(x^i, y^\sigma_I)$, with 
$
y^\sigma_I(j^r_x\gamma)=\frac{\partial^s \gamma^\sigma(x)}{\partial x^{i_1}\dots \partial x^{i_s}} |_x$, for $I=(i_1 \dots i_s)$ and $s\leq k$, is a system of coordinates on $J^rY$.

In the following we will always start a sum over multi-indices from the $0$-length multi-index, unless otherwise specified; the upper limit in such a sum will be usually given by multi-indices of length equal to the order of the jet prolongation under consideration. In these kind of sums, one should in principle restrict to {\em $s$-uple} of indices such that $i_1 \leq \dots \leq i_s$; however, multiplying the operators $\frac{\partial}{\partial y^\sigma_{I}}$ by suitable numerical factors, one is allowed to sum over all multi-indices $I$. 

Let $\Omega_q(J^kY)$ denote the module of $q$-forms on $J^kY$.
A major r\^ole in the calculus of variation is played by the so called contact structure induced by the affine bundle structure of $\pi_{k,k-1}$ (see \cite{Sau89}, \cite{Kru15}).  
A differential $q$-form $\alpha$ on $J^kY$ is called a contact form if, for every section $\gamma$ of $\pi$, we have 
\beq
 \left(j^k\gamma\right)^*(\alpha) = 0 \,.
\eeq 
It is easy to see that forms $\omega$  locally given as
\beq 
\omega^\sigma_{j_1\dots j_h}= dy^\sigma_{j_1\dots j_h}-y^\sigma_{j_1\dots j_h i}dx^i
\eeq 
for $0\leq h <k$ are indeed  contact $1$-forms.  In particular, it is easy to show that $(dx^i,\omega^\sigma,\omega^\sigma_{j_1},\dots, \omega^\sigma_{j_1\dots j_{k-1}}, dy^\sigma_{j_1 \dots j_k})$ is an alternative local basis for $1$-forms on $J^kY$. 
It is important to notice that the ideal of the exterior algebra generated by contact  forms on a fixed jet order prolongation is not closed under exterior derivation, while if $\alpha$ is contact so is $d\alpha$.

A \emph{Lagrangian of order $r$} is a horizontal $n$-form $\lambda$ on $J^rY$. Notice that any $q$-form $\rho\in \Omega_q(J^kY)$ can be written in a unique way as 
\[
\pi_{k+1, k}^*\rho=\sum\limits^q_{i=0} p_i\rho
\]
where $p_i\rho$ is generated by wedge products containing exactly $i$ factors of the type $\omega^\sigma_{j_1\dots j_l}$ for $0\leq l\leq k$; $\rho$ is contact if and only $p_0\rho=0$ and it is a Lagrangian if and only if it has degree $n$ and $p_i\rho=0$ for all $i>0$. We will use the notation $h\rho$ in place of $p_0\rho$. The form $h\rho$ is a Lagrangian of order $r+1$ and it is called the \emph{Lagrangian associated to $\rho$}. A form $\rho$ of degree $q>n$ is called \emph{strongly contact} if $p_{q-n}\rho=0$.
We finally introduce the \emph{horizontal differential} $d_H$, defined by
$
d_H\rho=\sum\limits^q_{i=0} p_{i} d p_i\rho
$,
and the \emph{vertical differential} $d_V$, defined by
$
d_V\rho=\sum\limits^q_{i=0} p_{i+1} d p_i\rho
$,
 such that
$
\pi_{k+2, k}^*d\rho=d_H\rho+d_V\rho$.

\subsection{Geometric variation formulae}

In the following we shall define the {\em interior Euler operator}  $\mathcal{I}$, which was introduced to the calculus of variations within the variational bicomplex theory \cite{An}
and adapted to the finite order situation of the variational sequence in \cite{Kr02,KrMu03,KrMu05,VoUr14}; see also the review in \cite{Kru15} and applications to the representation of variational Lie derivatives in \cite{CaPaWi16,PaRoWiMu16}.

We refer to so-called {\em contraction Euler operators}, \ie  Euler operators corresponding to the formal differential operator defined by the contraction
$J^{r+1}\Xi \rfloor p_k\rho $, where $\rho$ is a (local) $(n+k)$-form. 
Locally
\beq
p_k\rho=\sum\limits^{}_{0\leq |J_1|, \dots |J_k|\leq r}\rho^{J_1 \dots J_k}_{\sigma_1 \dots \sigma_k}\omega^{\sigma_1}_{J_1}\wedge \dots \wedge \omega^{\sigma_k}_{J_k}\wedge ds \,,
\eeq
and then
\beq
J^{r+1}\Xi\rfloor p_k\rho=\sum\limits^{r}_{|J|=0}d_J\Xi^\sigma\left(\frac{\partial}{\partial y^\sigma_J}\rfloor p_k\rho\right) \,.
\eeq
The corresponding contraction Euler operator $I$ is given locally by
$ I(\Xi)=\Xi^\sigma I_\sigma
$, where
\beq 
I_\sigma=\sum\limits^{r}_{|J|=0}(-1)^{|J|}d_J\left(\frac{\partial}{\partial y^\sigma_J}\rfloor p_k\rho\right) \,.
\eeq 

By means of this operator we can thus define a map
$\mathcal{I}:\Omega^r_{n+k}W\to \Omega^{2r+1}_{n+k}W$
by 
\begin{equation}
\mathcal{I}(\rho)=\frac{1}{k}\omega^\sigma \wedge I_\sigma=\frac{1}{k}\omega^\sigma\wedge\sum^{r}_{|I|=0}(-1)^{|I|}d_I\left(\frac{\partial}{\partial y^\sigma_I}\rfloor p_k\rho\right) \,.
\end{equation}
This map is called {\em interior Euler mapping} or {\em interior Euler operator}; it is a definition adapted to finite order jets from the one given by Anderson, see \cite{An}.
It turns out that, if $\rho$ is global,
 $\mathcal{I}(\rho)$ is a globally defined form. This operator can be defined intrinsically;  see \cite{KrMu03,KrMu05} and \cite{PaRoWiMu16}.

A local operator $\mathcal{R}$, called the {\em residual operator}, is defined by
\beq 
(\pi_{2r+1,r+1})^*(p_k\rho)=\mathcal{I}(\rho)+p_kdp_k\mathcal{R}(\rho) \,,
\eeq 
and $\mathcal{R}(\rho)$ is a local strongly contact ($n+k-1$)-form.

In the following, for any $n$-form $\rho$, 
we will use the notation
$E_n(h\rho) \doteq \cI (d \rho) =\mathcal{I}(dh\rho)$. As discussed in \cite{PaRoWiMu16} (see also the references therein), it turns out that $E_n(h\rho)$ is the Euler--Lagrange form obtained as the representation by the interior Euler operator of the variational class defined by $d \rho$ modulo a suitably defined contact structure having a meaning from the point of view of the calculus of variations \cite{Kru90}.

\bTh \label{var_lie_der} 
For  any $n$-form $\rho$ and for any $\pi$-projectable vector field $\Xi$ on $\bY$, we have, up to pull-backs by projections,
\begin{equation}\label{eq:var_n}
L_{J^{r+1}\Xi}h\rho=\Xi_V\rfloor E_n(h\rho)+ 
d_H(J^{r+1}\Xi_V\rfloor p_{d_Vh\rho} + \Xi_H\rfloor h\rho)  \,
\end{equation}
where
\beq p_{d_Vh\rho} = - p_1\mathcal{R}(dh\rho) \,.
\eeq 
\eTh
We stress that \eqref{eq:var_n} can be regarded as the local first variation formula for the Lagrangian $h\rho$ with respect to a (variation) projectable vector field; we refer the reader to \cite{Kru73,Kru15} for details. 
\bDf
The Noether current \cite{Noe18,Kos11} for a Lagrangian $\lambda$ associated with $\Xi$ is defined as 
\beq
\epsilon_{\Xi}(\lambda)=J^{r+1}\Xi_V\rfloor p_{d_V\lambda} + \Xi_H\rfloor \lambda\,.
\eeq
The term $p_{d_V\lambda} = - p_1\mathcal{R}(d\lambda)$ is called {\em a local generalized momentum}.
\eDf
Based on an iteration of the first variation formula expressed through interior Euler operator as above, a formula for the second variation of a Lagrangian can be obtained (see \cite{Ac17,AcPa17} for the extension to any variation order); it will be further explored in the next section. 

The idea is the following: we note that $L_{J^{r+1}\Xi}h\rho=hL_{J^r\Xi}\rho$, and then apply a standard inductive reasoning. Of course, the iterated variation is pulled-back up to a suitable order, in order to suitably split the Lie derivatives. 

\bTh\label{2nd_var}
Let $\rho$ be an $n$-form on $J^rY$, $[\rho]$ its class and $\lambda=h\rho$ the associated Lagrangian. For any pair of projectable vector fields $\Xi_1$ and $\Xi_2$, we have, up to pull-backs by projections,
\begin{equation}\label{eq:2nd_var}
\begin{aligned}
\left(L_{J^{r+1}\Xi_2}\right.&\left.L_{J^{r+1}\Xi_1}h\rho\right)=\Xi_{2,V}\rfloor E_n(\Xi_{1,V}\rfloor E_n(h\rho)) + \\
&+d_H\epsilon_{\Xi_2}(\Xi_{1,V}\rfloor E_n(h\rho))+d_H\epsilon_{\Xi_2}(d_H\epsilon_{\Xi_1}(h\rho))
\end{aligned}
\end{equation}
where the two following  Noether currents are defined
\begin{align*}
\epsilon_{\Xi_2}(\Xi_{1,V}\rfloor E_n(h\rho))=&\Xi_{2,H}\rfloor \Xi_{1,V}\rfloor E_n(h\rho) +\\
&+J^{r+1}\Xi_{2,V}\rfloor p_{d_V \Xi_{1,V}\rfloor E_n(h\rho)}  \,, \\
\epsilon_{\Xi_2}(d_H\epsilon_{\Xi_1}(h\rho))=&\Xi_{2,H}\rfloor d_H(J^{r+1}\Xi_{1,V}\rfloor p_{d_Vh\rho}+\Xi_{1,H}\rfloor h\rho)+\\
&+J^{r+1}\Xi_{2,V}\rfloor p_{d_V d_H (J^{r+1}\Xi_{1,V}\rfloor p_{d_V h\rho}+\Xi_{1,H}\rfloor h\rho)} \,.
\end{align*}
\eTh

\bRm {\em
Some useful identities follow by the above Theorem and by the naturality of the Lie derivative. 
In particular,
for every pair of {\em  vertical vector fields} $\Xi_1$ and $\Xi_2$, we have
\bEq\label{eq:2nd_commu}
&d_H(\epsilon_{\Xi_2}(\Xi_{1}\rfloor E_n(h\rho)))=
\\ 
&\Xi_{1}\rfloor E_n (\Xi_{2}\rfloor E_n(h\rho)) - \Xi_{2}\rfloor E_n(\Xi_{1}\rfloor E_n(h\rho)) + [\Xi_2,\Xi_{1}]\rfloor E_n(h\rho)\,.\nonumber 
\eEq
Furthermore, we have
$d_H [ j^{k+1}\Xi_{2,V}\rfloor p_{d_V d_H (\epsilon_{\Xi_1}(h\rho))} - h( \Xi_{2,V}\rfloor d_V(j^{k+1}\Xi_{1,V}\rfloor p_{d_Vh\rho})-\Xi_{1,H}\rfloor h\rho)]=0$, and 
$
d_H[\epsilon_{\Xi_2}(\Xi_{1}\rfloor E_n(h\rho) +\epsilon_{\Xi_1}(\Xi_{2}\rfloor E_n(h\rho))]=0$.}
\eRm

\section{The Jacobi morphism} \label{Jacobisection}

Here the definition of the Jacobi morphism will be given by the interior Euler operator;  for slightly different approaches and further applications see \eg \cite{FFPW08,FFPW11,FPV05,FrPaWi05,PaWi04,PaWi03,PaWi07,PaWi08b,PaWi11,PaWi14,PaWi19}. We recover that the Jacobi morphism is self adjoint along extremals, finding also explicit coordinate expressions, and we introduce the Jacobi equation and Jacobi fields. In this framework, we easily define  the Hessian of the action, which turns out to be related to the Jacobi morphism. Our discussion is inspired by the classical paper \cite{GoSt73}, where, for  first order field theories, Goldschmidt and Sternberg gave the definitions of {\em Hessian} and {\em Jacobi equation}.

We recall 
the expression of the
adjoint of a differential operator associated with a suitable $(n+2)$-form. 
Consider a global $(n+2)$-form on $J^{r+1}Y$ with local coordinate expression given by
\beq 
\omega=\sum\limits^{r}_{|J|=0}A^{J}_{\tau\sigma}\omega^{\tau}_{J}\wedge \omega^\sigma\wedge ds \,;
\eeq 
the local expressions for  $\mathcal{I}(\omega)$ are
\beq 
\mathcal{I}(\omega)=\sum\limits^{k}_{|J|=0}\frac{1}{2}\omega^\tau \
\wedge  (-1)^{|J|}d_J(A^J_{\tau\sigma} \omega^\sigma)\wedge ds-\sum\limits^{k}_{|J|=0}\frac{1}{2}\omega^\sigma\wedge A^{J}_{\tau\sigma}\omega^{\tau}_{J}\wedge ds \,.
\eeq 
We can now introduce $\tilde{\mathcal{I}}(\omega)$, associated with $\mathcal{I}(\omega)$ and defined as
\beq 
\tilde{\mathcal{I}}(\omega)=-\sum\limits^{r}_{|J|=0}(-1)^{|J|}d_J(A^J_{\rho\sigma} \omega^\sigma)\otimes\omega^\rho \otimes   ds+\sum\limits^{r}_{|J|=0}A^{J}_{\tau\sigma}\omega^{\tau}_{J}\otimes\omega^\sigma\otimes  ds \,.
\eeq 
Let us set
\beq 
\tilde{\omega}=\sum\limits^{r}_{|J|=0}A^{J}_{\tau\sigma}\omega^{\tau}_{J}\otimes\omega^\sigma\otimes ds \,.
\eeq 
We introduce some formal differential operators associated with $\omega$. Define
\beq
\nabla_\omega :\ X_V(Y)& \to C^1_0 \otimes \Omega^r_{n,X}(J^rY)\\
\Xi& \mto  \tilde{\omega}  (J^{r+1}\Xi, \bullet ) \,
\eeq
where we have denoted by $X_V(Y)$ the space of vertical vector fields on $Y$ and by $C^1_0$ the space of contact $1$-forms generated by $\omega^\sigma$. In coordinates 
\beq 
\nabla_\omega\left(\Xi^\sigma \frac{\partial}{\partial y^\sigma}\right)=\sum\limits^{r}_{|J|=0}A^{J}_{\tau\sigma}d_{J}(\Xi^{\tau})\omega^\sigma \otimes ds
\eeq 
Moreover we set
\beq
\nabla^*_\omega :\ X_V(Y)& \to C^1_0 \otimes \Omega^r_{n,X}(J^rY)\\
\Xi& \mto( \tilde{\omega}-\tilde{\mathcal{I}}(\omega))   ( J^{r+1}\Xi, \bullet)
\eeq
that in coordinates is
\beq 
\nabla^*_\omega \left(\Xi^\sigma \frac{\partial}{\partial y^\sigma}\right)=\sum\limits^{r}_{|J|=0}(-1)^{|J|}d_J(A^J_{\tau\sigma} \Xi^\sigma)\omega^\tau \otimes   ds\,.
\eeq 
The choice in the notation is motivated by the fact that $\nabla^*_\omega$ can be seen as an {\em adjoint operator} for $\nabla_\omega$.

\bDf 
The map 
\begin{equation}\label{eq:Jacobi}
\begin{aligned}
\mathcal{J}:\Omega^r_{n,X}(J^rY)&\to X^*_V(J^{2r+1}Y)\otimes X^*_V(Y)\otimes \Omega^r_{n,X}(J^rY)\\
\lambda &\mto \bullet\ \rfloor E_n( \bullet\ \rfloor E_n(\lambda)) 
\end{aligned}
\end{equation}
is called the Jacobi morphism associated with $\lambda$. In the following we use the notation
$ \mathcal{J}_{\Xi_1}(\lambda)\doteq E_n(\Xi_1\rfloor E_n(\lambda))$.
\eDf

In the following theorem we state an important property of the Jacobi morphism along solutions of the Euler--Lagrange equations (critical sections or extremals). We note that this property has been pointed out in \cite{GoSt73} for first order field theories and has been extended to higher order field theories in \cite{FPV05,PaWi07} by different approaches, both referring to Kola\v r's decomposition formulae of vertical morphisms \cite{Kol83}.

Following a different approach, by means of Krbek-Musilov\'a's geometric integration by parts and by the exactness of the variational sequence, we prove straightforwardly that the Jacobi morphism is self-adjoint along critical sections of a Lagrangian field theory of any order (see also \cite{AcPa19,FPV05}). 
This is a property of great importance in physical applications.

\bTh\label{th:Jacobi}
For any pair  of vertical  vector fields $\Xi_1$, $\Xi_2$ on $Y$, we have
\beq 
J^{2r+1}\Xi_2\rfloor\mathcal{I}(J^{2r+1}\Xi_1\rfloor d \mathcal{I}(d\lambda))=0\,.
\eeq 
Along extremals the Jacobi morphism is self adjoint. 
\eTh

\bPf
Since $\lambda=p_0\lambda$, up to pull-backs, 
\beq 
0 = d d\lambda =  dp_1d\lambda = d\mathcal{I}(d\lambda) +d p_1dp_1\mathcal{R}(d\lambda) \,,
\eeq 
holds true and we have that
$ -  J^{2r+1}\Xi\rfloor  d\mathcal{I}(d\lambda)$ is an $n+1$ contact form for every vertical vector field $\Xi$; this implies that 
\beq 
J^{2r+1}\Xi_2\rfloor\mathcal{I}(J^{2r+1}\Xi_1\rfloor d \mathcal{I}(d\lambda))=0\,,
\eeq 
for any pair of vertical vector fields $\Xi_1$, $\Xi_2$.

Let $E_\rho(\lambda)$ be the local components of the Euler--Lagrange form associated with $\lambda$. We therefore have the local condition
\beq
\mathcal{I}(J^{2r+1}\Xi\rfloor d \mathcal{I}(d\lambda)) & = &  \sum\limits^{2r+1}_{|J|=0}\frac{\partial E_\sigma(\lambda)}{\partial y^\rho_J}d_J\Xi^\rho \omega^\sigma \wedge ds +\\
&-& \sum\limits^{2r+1}_{|J|=0}(-1)^{|J|}d_J(\frac{\partial E_\sigma(\lambda)}{\partial y^\rho_J}\Xi^\sigma)\omega^\rho\wedge ds =0 \,,
\eeq

Note that {\em along extremals} the terms of the form 
$\frac{\partial \Xi^\rho}{\partial y^\sigma}E_\rho(\lambda)$ 
vanish. Therefore, for every vertical vector field $\Xi$ on $Y$, we have the equality of the following two local expressions (the first coming from the direct calculation of $E_n(\Xi\rfloor E_n(\lambda))$ along extremals, the second coming from the identity above):
\bEq\label{eq:coordJacobi2}
E_n(\Xi\rfloor E_n(\lambda))=\sum\limits^{2r+1}_{|J|=0}(-1)^{|J|}d_J(\Xi^\rho\frac{\partial E_\rho(\lambda)}{\partial y^\sigma_J})\omega^\sigma \wedge ds
 =
\eEq
\bEq\label{eq:coordJacobi}
 =\sum\limits^{2r+1}_{|J|=0} d_J\Xi^\sigma\frac{\partial E_\rho (\lambda)}{\partial y^\sigma_J}\omega^\rho\wedge ds 
\,.
\eEq
These two local expressions provide, indeed, two (adjoint to each other) expressions for the Jacobi morphism along extremals,  which is then self-adjoint.
\ePf
Note that, of course, $\cI d(\Xi\rfloor \cI d(\lambda)) \equiv E_n(\Xi\rfloor E_n(\lambda))$ should not be confused with $\mathcal{I}(J^{2r+1}\Xi\rfloor d \mathcal{I}(d\lambda))$, which is instead related to the Helmholtz form expressing conditions of local variationality for a source form.

\bDf
Let $\lambda$ be a Lagrangian of order $r$. A {\em Jacobi field} for the Lagrangian $\lambda$ is a vertical vector field $\Xi$ that belongs to the kernel of the Jacobi morphism, {\em i.e.}
\beq \mathcal{J}_{\Xi}(\lambda)=0
\eeq 
The equation above is called  the {\em Jacobi equation} for the Lagrangian $\lambda$.
\eDf
The Jacobi equation
evaluated along an extremal $\gamma$  depends only on the values of the vector field $\Xi$ along $\gamma$;
 we therefore correctly can speak of {\em Jacobi fields along an extremal} $\gamma$.
 
 Note that  Theorem \ref{th:Jacobi} provides us with the coordinate expressions of the equation for Jacobi fields along an extremal.

\subsection{The Jacobi morphism and the Hessian of the action functional}

We reformulate, within our formalism, some results due to Goldschmidt and Sternberg \cite{GoSt73}, which enable us to generalize to any arbitrary order; in particular by the above definition of a Jacobi morphism we recover the definition of the  Hessian  and generalize some of its properties at any order (see Propositions \ref{H1}  and \ref{H2}). 

In the following we say that a vector field on $J^rY$, with $r\geq 0$, vanishes identically on a subset of $X$, meaning that it vanishes along the fibers over that subset.
We introduce a suitable notion of variation by which we handle different aspects of the theory in a unified way.
First, we shall define variations of sections.

\bDf
Consider an open subset $U\subseteq X$ and a section $\gamma$ of $\pi$ defined on $U$. Take an integer $s\geq 0$. An {\em $s$-parameters variation of $\gamma$} is a map $\Gamma: I^s\times X\to Y$, where $I^s$ denotes the $s$-cube with side $]-1,1[$, such that
\begin{enumerate}
\item[-] if $i_X$ denote the inclusion $X \to I^s\times X$, then $\Gamma\circ \left.i_X\right|_U : U \to Y$ is a section;
\item[-] we have $\Gamma(0,\dots,0,x)=\gamma(x)$ for any $x \in U$.
\end{enumerate}
\eDf
Let now $\Xi_1, \dots, \Xi_s$ be vertical vector fields on $Y$ such that
\begin{align*}
\left.\frac{\partial \Gamma(t_1, 0, \dots, 0)}{\partial t_1}\right|_{t_1=0}&=\Xi_1\circ \gamma\\
\left.\frac{\partial \Gamma(t_1, t_2, 0, \dots, 0)}{\partial t_2}\right|_{t_2=0}&=\Xi_2 \circ \Gamma(t_1, 0 \dots 0,x)\\
\dots& \\
\left.\frac{\partial \Gamma(t_1,\dots t_s)}{\partial t_s}\right|_{t_s=0}&=\Xi_s\circ \Gamma(t_1, t_2, \dots t_{s-1},0, x) \,.
\end{align*}
We say that $\Gamma$ is {\em generated by the variation vector fields $\Xi_1, \dots, \Xi_s$} and write $\Gamma_{\Xi_1, \dots \Xi_s}$. 
This definition holds true also for projectable vector fields; see \cite{Kru15}.

This enables us to suitably define variations of forms along sections.
\bDf
Let $\rho\in \Omega^k_qW$ be a local $q$-form. Consider an $s$-parameters variation $\Gamma$ of a section $\gamma$. The {\em $s$-variation of $\rho$ along $\gamma$ associated with $\Gamma$}, denoted by $\delta^s_{\Gamma}\rho$, is
\beq 
\left.\delta^s_{\Gamma}\rho\right|_{j^r\gamma(x)}=\left.\frac{\partial^s(\rho \circ j^r\Gamma(t_1, \dots ,t_s, x))}{\partial t^1 \dots \partial t^s}\right|_{t_1=\dots=t_s=0} \,.
\eeq 
\eDf
Variations of forms are Lie derivatives; indeed the following holds true.
\bPr
Let $\rho\in \Omega^k_qW$ be a local $q$-form and consider an $s$-variation of $\rho$ along $\gamma$ associated with $\Gamma$, where $\Gamma$ is generated by $\Xi_1, \dots, \Xi_s$. Then
\beq 
\left.\delta^s_\Gamma \rho\right|_{\gamma(x)}=L_{J^r\Xi_1} \dots L_{j^r\Xi_s}\rho \circ j^r\gamma(x) \,.
\eeq 
\ePr
Moreover, {\em formal variations generated by $\Xi_1, \dots, \Xi_s$}, for short {\em variations} - see \cite{FPV05}, are defined as 
\beq 
\delta_{\Xi_1, \dots, \Xi_s}\rho=L_{J^r\Xi_1} \dots L_{j^r\Xi_s}\rho \,.
\eeq 

\bDf
Let $\Gamma$ be a one parameter variation of a section $\gamma$ and let $\Xi$ be the variation vector field; $t$ will denote the parameter of the flow of $\Xi$. The {\em variation of the action induced by $\Xi$ and evaluated at $\gamma$} is defined as
\beq 
\delta_\Xi A_D[\gamma]=\left.\frac{d}{dt}\right|_{t=0}A_D[\psi_t\circ \gamma ]=\left.\frac{d}{dt}\right|_{t=0}\int\limits_{D}\lambda \circ j^r\psi_t \circ j^r\gamma 
\eeq 
\eDf
The variation of the action can be expressed as
\beq
\int\limits_{D}(j^r\gamma)^*(L_{J^r\Xi}\lambda)=\int\limits_{D}L_{J^r\Xi}\lambda\circ j^r\gamma \,.
\eeq Note that the above formula holds true also for projectable vector fields (see, \eg \cite{Kru15}).
We can easily generalize it to iterated variations. 

The $s$-th variation of the action generated by $\Xi_1, \dots, \Xi_s$ at $\gamma$ is given by 
\beq
\delta_{\Xi_1, \dots \Xi_s} A_D[\gamma]=\int\limits_D L_{J^r\Xi_1} \dots L_{J^r\Xi_s} \lambda \circ j^r\gamma \,.
\eeq
Finally we can give the definition of extremal. If $\Xi$ is a vector field along a local section $\gamma$ defined on $U$, $\overline{\Xi}$ denote the set
\beq 
\text{cl}\{x \in U \text{ s.t. } \Xi|_x\neq 0\} \,.
\eeq 
where \text{cl} means closure. Recall that it is possible to extend $\Xi$ to a vector field $\tilde{\Xi}$ defined in a neighborhood of $\gamma(U)$. 
\bDf
A section $\gamma$ defined on $U$ is called {\em extremal} of the action functional (or of the action) if, for every $\pi$-vertical vector fields $\Xi$ such that $\overline{\Xi \circ \gamma}\subseteq D$, it holds
\beq 
\delta_\Xi A_D(\gamma)=\int \limits_{D} L_{J^r\Xi} \lambda \circ j^r\gamma=0 \,.
\eeq 
\eDf

Take the action functional associated with a Lagrangian $\lambda$ on $J^rY$:
\beq A_D[\gamma]=\int\limits_{D} (j^r\gamma)^*(\lambda)
\eeq 
where $D$ is an $n$-region and $\gamma$ is a section. We can introduce variations that do not change $\gamma$ on $\partial D$; it is sufficient to generate it with vector fields that vanish on $\partial D$. We will work only on $D$ so we can require that these fields vanish also in $X\backslash D$ (all vector fields compactly supported in an open proper subset of $D$ are an example). 

We denote by $\Gamma^\tau_{D}$ the space of sections defined on $D$ and equal to a fixed section $\tau$ on $\partial D$, by $X_{V,\gamma}(Y)$ the space of vertical vector fields defined along $\gamma\in \Gamma^\tau_{D}$ and by $T_\gamma\Gamma^\tau_{D}$ the subspace of $X_{V,\gamma}(Y)$ containing vector fields that are null on $\partial D$. The last notation is motivated by the fact that $T_\gamma\Gamma^\tau_{D}$ can be thougt as the tangent space to $\Gamma^\tau_{D}$ at the point $\gamma$ (this is a quite standard fact from the theory of infinite dimensional manifolds). We then give the following definition (which is standard too).
\bDf
The differential of $A_D$ along a section $\gamma\in \Gamma^\tau_{D}$ is the map
\begin{equation}\label{eq:diff}
\begin{aligned}
dA_D[\gamma]: T_\gamma \Gamma^\tau_{D}& \to \mathbb{R}\\
               \nu& \to  \left.\frac{d}{dt}A_D[\Gamma_\Xi(t)]\right|_{t=0}
\end{aligned}
\end{equation}
where $\Xi$ is any extension of $\nu$ to the whole $X_V(Y)$ vanishing on $\partial D$ (and on $X/D$), while $\Gamma_\Xi$ is a one parameter variation of $\gamma$ generated by $\Xi$.
\eDf
In our hypothesis the derivation passes under the integral and we have
\beq dA_D[\gamma](\nu)=\int\limits_{D}(j^r\gamma)^*(\delta_\Xi\lambda) \,.
\eeq 
We need to show that the definition does not depend on $\Xi$, but thanks to this last remarks this means that we need to prove
\beq \int\limits_{D}(j^r\gamma)^*(L_{J^r\Xi}\lambda)=\int\limits_{D}(j^k\gamma)^*(L_{J^r\Xi'}\lambda) \,,
\eeq 
where $\Xi'$ is an alternative choice of the field. Here we simply apply \eqref{eq:var_n} and note that the horizontal differential plays no r\^ole thanks to Stokes theorem and the fact that the fields vanish on $\partial D$; then we use the fact that, because of the pull-back by $j^r\gamma$, everything depends only on the values of the fields along $\gamma$.

So we have a notion of differential of the action; an extremal point is exactly a solution of the Euler--Lagrange equations. In an analogous manner, we introduce an Hessian for the action, but we need the section $\gamma$ to be an extremal. 
\bDf
The Hessian of $A_D$ along an extremal section $\gamma \in \Gamma^\tau_D$ is the map
\begin{equation}\label{eq:hessian}
\begin{aligned}
\mathcal{H}(A_D)[\gamma]:T_\gamma \Gamma^\tau_D \times T_\gamma\Gamma^\tau_D & \to \mathbb{R}\\
(\nu, \kappa)&\to \left.\frac{\partial^2}{\partial t_1\partial t_2}A_D[\Gamma_{\Xi_1,\Xi_2}(t_1,t_2)]\right|_{t_1=t_2=0}
\end{aligned}
\end{equation}
where $\Xi_1$ and $\Xi_2$ are extensions in $X_V(Y)$ of $\nu$ and $\kappa$ respectively and vanish on $\partial D \cup X\backslash D$, while $\Gamma_{\Xi_1,\Xi_2}(t_1,t_2)$ is a two parameter variation of $\gamma$ generated by $\Xi_1$ and $\Xi_2$.
\eDf
Deriving again under the integral sign
\beq \mathcal{H}(A_D)[\gamma](\nu,\kappa)=\int\limits_D (j^r\gamma)^*(L_{J^r\Xi_1}L_{J^r\Xi_2}\lambda) \,.
\eeq 
In order to show that the Hessian is well defined, we need to prove that it does not depend on the extensions $\Xi_1$ and $\Xi_2$ chosen; moreover, we want to show that it is symmetric. Here we need the hypothesis that the section is an extremal; in fact this implies, by applying \eqref{eq:2nd_commu}, the Stokes theorem and the fact that the fields vanish on $\partial D$, 
\beq
 \int\limits_D (j^r\gamma)^*(L_{J^r\Xi_1}L_{J^r\Xi_2}\lambda)=\int\limits_D (j^r\gamma)^*(L_{J^r\Xi_2}L_{J^r\Xi_1}\lambda) \,.
\eeq 
Now, using \eqref{eq:2nd_var}, we see that the definition can be restated as
\beq \mathcal{H}(I_A)[\sigma](\nu,\kappa)=\int\limits_{D}(j^{2+1}\gamma)^*(\Xi_1 \rfloor E_n(\Xi_2\rfloor E_n(\lambda))) \,,
\eeq 
that obviously does not depends on the extension $\Xi_1$. However, by symmetry it cannot depend on $\Xi_2$ too. Then the Hessian is a well defined symmetric bilinear map.

Now we can easily recover the properties stated at the first order by Goldschmidt and Sternberg in \cite{GoSt73}: we need only to apply our intrinsic results.
\bPr \label{H1}
If $\gamma$ is a local minimum, the Hessian along it is positive semi definite.
\ePr
\bPf
We consider $\Gamma_\Xi$, a one parameter variation of $\gamma$ generated by $\Xi$. We have by hypothesis
\beq 
\int\limits_D(j^r\Gamma_\Xi(t))^*(\lambda)\geq\int\limits_D(j^r\gamma)^*(\lambda) \,,
\eeq 
for all $t$ in a neighborhood of $0$. Consequently
\beq \frac{d^2}{dt^2}\int\limits_D (j^r\Gamma_\Xi(t))^*(\lambda)\geq 0 \,.
\eeq 
But, by definition, the left hand side is the Hessian along $\gamma$ calculated on the pair $(\Xi\circ \gamma, \Xi\circ \gamma)$. Thanks to the arbitrariness of $\Xi$ we can conclude that the quadratic form associated with the Hessian along $\gamma$ has only values $\geq 0$, which is our claim.
\ePf

Our construction gives us immediately also the relation between the Hessian and the Jacobi morphism. In fact it is clear from the previous discussion that
\beq 
\mathcal{H}(A_D)[\gamma](\nu,\kappa)=\int\limits_{D}(j^{2r+1}\gamma)^*(\Xi_1\rfloor \mathcal{J}_{\Xi_2 }(\lambda)))\,.
\eeq 
Finally, considering that our definitions of Hessian and Jacobi field do not depend on the order, we can prove the following result applying the same argument used in \cite{GoSt73} for the first order case.

\bPr \label{H2}
A vector field $\nu \in T_\gamma\Gamma^\tau_D$, where $\gamma$ is an extremal, belongs to the null space of the Hessian along $\gamma$ if and only if it is a Jacobi field along $\gamma$.
\ePr
\bPf
If we have a Jacobi field, then it is immediate to see that it belongs to the null space of the Hessian, because of the relation between the Hessian and the Jacobi morphism. 

Conversely, if $\nu \in T_\gamma\Gamma^\tau_D$ is in the nullspace of the Hessian, then
\beq \int\limits_D(j^{2r+1}\gamma)^*(\Xi_1 \rfloor \mathcal{J}_{\Xi_2}(\lambda))=0 \,,
\eeq 
for any $\Xi_1$ and any $\Xi_2$ that extends $\nu$. The arbitrariness of $\Xi_1$ enables us to conclude.
\ePf

\subsection{Symmetries of variations and Jacobi fields}

By using the approach focused on iterated variational Lie derivatives, we can investigate the existence of conservation laws associated with the Jacobi equation. 
In particular we investigate how conservation laws are associated to Jacobi fields and symmetries of variations.

\bDf
Let $\lambda$ be a Lagrangian on $J^rY$ (or on an open subset $V^r\subseteq J^rY$ where $V$ is open in $Y$). A {\em symmetry of $\lambda$} is an automorphism $f$ of $Y$ such that $J^rf$ is an invariance transformation of $\lambda$. 
\eDf 
By abuse of notation, we will use the term {\em symmetries} for infinitesimal generators of symmetries too.

\bDf A projectable vector field $\Xi$ is a symmetry of $\lambda$ if and only if 
\beq 
L_{J^r\Xi}\lambda=0 \,.
\eeq 
\eDf

Symmetries of a Lagrangian constitute a subalgebra of the algebra of vector fields on $J^rY$.

\bDf
 A projectable vector field $\Xi$ is a generator of invariance transformations for  a source form  $\omega \in \Omega^r_{n+1,Y}V$ if and only if 
\beq 
L_{J^r\Xi}\omega=0 \,.
\eeq 
\eDf
 Generators of invariance transformations constitute a subalgebra of the algebra of projectable vector field on $J^rY$.
 An invariance transformation of $\lambda$ is an invariance transformation of $E_\lambda$; furthermore
given an invariance transformation $f$ of $E_\lambda$, $\lambda-(J^rf)^*(\lambda)$ is a trivial Lagrangian; see \eg \cite{Kru15,Tra67}.

An invariance transformation of $E_\lambda$ is called {\em generalized symmetry} of $\lambda$. A generator of invariance transformations of $E_\lambda$ is an {\em infinitesimal generator of generalized symmetries} of $\lambda$; we will call it simply {\em generalized symmetry}.

\bDf
Given a section $\gamma\in \Gamma_{loc}(\pi)$ and an open set $W$ in $J^rY$, an $(n-1)$-form $\epsilon\in \Omega^r_{n-1,X}W$ such that
\beq d (j^r\gamma)^*(\epsilon)=(j^{r+1}\gamma)^*d_H\epsilon=0 \,,
\eeq 
is called {\em conserved current} along $\gamma$. The previous equality is a {\em weak conservation law} along $\gamma$.
\eDf
The term {\em weak} is related to the fact that the form $\epsilon$ is closed (equivalently, {\em horizontally closed}) only along the section $\gamma$. When  a current    is {\em horizontally closed} everywhere, one speaks of a {\em strongly} conserved current. 

Our results concerning the relation between the second variation and the Jacobi fields can be applied as follows.
\bTh
\label{Jacobicons}
Let $\rho$ be an $n$-form on $J^{r-1}Y$ and $h\rho$ the associated Lagrangian on $J^rY$. Consider two vertical vector fields $\Xi_1$ and $\Xi_2$ on $Y$ . 
\begin{enumerate}
\item Suppose that $\Xi_2$ is a symmetry of the first variation of $h\rho$ generated by $\Xi_1$ and that $\Xi_1$ and $\Xi_2$ satisfy
\beq \Xi_2\rfloor\mathcal{J}_{\Xi_1}(h\rho)=0 \,,
\eeq 
then
\begin{equation}\label{eq:Jacobicons1}
d_H\epsilon_{\Xi_2}(L_{J^{r}\Xi_1 }h\rho)=0 \,.
\end{equation}
\item Suppose that $\Xi_1$ and $\Xi_2$ are Jacobi fields, \ie 
\beq \mathcal{J}_{\Xi_i}(h\rho)=0 \,,
\eeq 
then, along
critical sections of $h\rho$,
\begin{equation}\label{eq:Jacobicons2}
d_H\epsilon_{\Xi_2}(\Xi_1\rfloor E_n(h\rho))=0 \,.
\end{equation}
\end{enumerate}
\eTh

\bRm {\em
We stress that \eqref{eq:Jacobicons1} can be interpreted as a strong conservation law. On the other hand \eqref{eq:Jacobicons2} can be seen as a weak conservation law associated with Jacobi fields. We can conclude that, taking two Jacobi fields $\Xi_1$ and $\Xi_2$ and working along
critical solutions, $d_H\epsilon_{\Xi_2}(\Xi_1\rfloor E_n(h\rho))
$ and $d_H\epsilon_{\Xi_2}(d_H\epsilon_{\Xi_1}(h\rho))
$ vanish separately.}
\eRm

\bRm {\em
We note that if the hypothesis of Theorem \ref{Jacobicons} hold along
critical solutions, then \eqref{eq:Jacobicons1} and \eqref{eq:Jacobicons2} hold along
critical solutions too, as we can see in a completely analogous manner. }
\eRm

\section{The Jacobi equation for a Yang-Mills theory on a Minkowskian background} \label{Yang--Mills}

In this section we discuss the Jacobi equation for Yang-Mills theories \cite{YaMi54} on a Minkowskian background. We present a short separate discussion of the Maxwell case, then we work out in detail
the result for the Yang-Mills case, which is given by Equation \eqref{JACOBI EQUATION}; this equation has been obtained in \cite{Ac17} and it plays an important r\^ole within the invariance theory of the set extremals \cite{AcPa19} as well as within a variational characterization of Higgs fields \cite{PaWi19}. 

Note that in this example, the (configuration) fibered manifolds have the structure of bundles.
We work out an explicit computation by using the interior Euler operator,
 and we will refer to it later in relation with conservation laws; see subsection \ref{Jac_conservation}. 
 
In the sequel, by assuming a {\em Minkowskian background} we mean that the {\em spacetime manifold}, that is the base space $M$ of the {\em configuration bundle} for the theory, is equipped with a fixed {\em Minkowskian metric}, 
that is a {\em flat Lorentzian metric}. 
This means that $M$ is a {\em Lorentzian manifold} such that we can choose a system of coordinates in which the metric is expressed in the {\em diagonal form} $\eta_{\mu \nu}$; there are two possible conventions on the signature but, since the signature will play no r\^ole in our discussion, the choice is left to the reader.

Consider a principal bundle with a {\em semi-simple structure group} $G$, denoted by $(P, p, M, G)$; this bundle is the {\em structure bundle} of the theory. We consider the bundle $(C_P, \pi ,M)$ of {\em principal connections} on $P$ (the total space can be seen as $J^1 P \slash G$; see \cite{Sau89}). In this section lower Greek indices will denote space time indices, while capital Latin indices label the Lie algebra $\mathfrak{g}$ of $G$. Then, on the bundle $C_P$, we introduce coordinates $(x^\mu, \omega^A_\sigma)$.

We define the {\em Yang-Mills Lagrangian} as follows. We consider the {\em Cartan-Killing metric} $\delta$ on the Lie algebra $\mathfrak{g}$, and choose a $\delta$-orthonormal basis $T_A$ in $\mathfrak{g}$; the components of $\delta$ will be denoted $\delta_{AB}$. We will use $\delta_{AB}$ to raise and lower Latin indices. 
On a spacetime manifold with a generic (not necessarily Minkowskian) fixed background $g_{\mu \nu}$ the Yang-Mills Lagrangian is defined by
\beq 
\lambda_{YM}=-\frac{1}{4}F^A_{\mu \nu}g^{\mu \rho}g^{\nu \sigma}F^B_{\rho \sigma}\delta_{AB}\sqrt{g}ds \,,
\eeq 
where 
\begin{itemize}
\item $g$ stands for the absolute value of the determinant of the metric $g_{\mu \nu}$
\item we set $\omega^A_{\mu,\nu}=d_\nu\omega^A_\mu$;
\item we denote by $c^A_{BC}$ the {\em structure constants} of $\mathfrak{g}$;
\item $F^A_{\mu \nu}=\omega^A_{\nu,\mu}-\omega^A_{\mu,\nu}+c^A_{BC}\omega^B_\mu\omega^C_\nu$ is the so called  {\em field strength}.
\end{itemize}
The well known variation procedure 
for this Lagrangian gives the equations
\beq 
E^\nu_B 
= d_\mu(\sqrt{g}F^{\mu \nu}_B) +\sqrt{g}F^{\mu \nu}_A c^A_{BC}\omega^C_\mu=0 \,.
\eeq 
From now on, we will work with an atlas such that, in a coordinate chart, the metric $\eta$ is the standard Minkowskian metric.

We start our discussion from Maxwell theory; it is well known that we can regard it as an abelian Yang-Mills theory where the structure group is $U(1)$. Being the group one dimensional we can drop Latin indices; moreover, the structure constants vanish identically. We have then the well known equations:
\beq E^\nu=d_\mu(F^{\mu \nu})=0 \,.
\eeq 

Referring to theorem \ref{th:Jacobi}, we derive the expression of the Jacobi equation along 
critical solutions by  using the explicit formula \eqref{eq:coordJacobi}; the self adjointness property thus notably simplifies calculations.
Indeed, a vertical vector field $\Xi=\Xi_\sigma \frac{\partial}{\partial \omega_\sigma}$ satisfies  the Jacobi equation along
critical solutions if and only if
\beq
 \Xi_\sigma\frac{\partial E^\nu}{\partial \omega_\sigma}+ d_\rho \Xi_\sigma \frac{\partial E^\nu}{\partial \omega_{\sigma,\rho}}+ d_\theta d_\rho \Xi_\sigma \frac{\partial E^\nu}{\partial \omega_{\sigma,\rho \theta}}=0 \,.
\eeq 
Clearly, in this particular case,
\beq 
\frac{\partial E^\nu}{\partial \omega_\sigma}=0 \text{ \ \ \ and\ \ \ \ }
\frac{\partial E^\nu}{\partial \omega_{\sigma,\rho}}=0 \,,
\eeq 
so that we need only to calculate the last term.
Note that
\beq 
d_\mu(F^{\mu \nu})=\eta^{\alpha \mu}\eta^{\beta \nu}(\omega_{\beta, \alpha \mu}-\omega_{\alpha, \beta \mu}) \,
\eeq 
and so, by commutativity of total derivatives we get easily that our system of equations is equivalent to have, for every $\nu$,
\begin{equation}\label{JacobiMaxwell}
\eta^{\sigma\nu}\eta^{\tau\kappa}d_\kappa(d_\tau \Xi_\sigma-d_\sigma \Xi_\tau)=0 \,.
\end{equation} 
If we consider any semi-simple group $G$ instead of $U(1)$, a vertical vector field has the form
\beq 
\Xi=\Xi^Z_\sigma \frac{\partial}{\partial \omega^Z_\sigma} \,;
\eeq 
 in the particular case of investigation, since 
 \beq 
d_\mu(F^{\mu \nu}_B)=\delta_{BA}\eta^{\lambda \mu}\eta^{\sigma \nu}(\omega^A_{\sigma, \lambda \mu}-\omega^A_{\lambda, \sigma \mu}+c^A_{CD}\omega^C_{\lambda,\mu}\omega^D_\sigma+c^A_{CD}\omega^C_\lambda\omega^D_{\sigma,\mu}) \,,
\eeq 
and
\beq 
F^{\mu \nu}_Ac^A_{BC}\omega^C_\mu=\eta^{\lambda\mu}\eta^{\sigma\nu}\delta_{DA}\left(\omega^D_{\sigma,\lambda}-\omega^D_{\lambda,\sigma}+c^D_{EF}\omega^E_\lambda\omega^F_\sigma\right)c^A_{BC}\omega^C_\mu \,,
\eeq 
the Euler--Lagrange expressions for the Yang--Mills Lagrangian
are given by
 \beq
E^\nu_B = \delta_{BA}\eta^{\lambda \mu}\eta^{\sigma \nu}(\omega^A_{\sigma, \lambda \mu}-\omega^A_{\lambda, \sigma \mu}+c^A_{CD}\omega^C_{\lambda,\mu}\omega^D_\sigma+c^A_{CD}\omega^C_\lambda\omega^D_{\sigma,\mu})
 + \\
 + \eta^{\lambda\mu}\eta^{\sigma\nu}\delta_{DA}\left(\omega^D_{\sigma,\lambda}-\omega^D_{\lambda,\sigma}+c^D_{EF}\omega^E_\lambda\omega^F_\sigma\right)c^A_{BC}\omega^C_\mu \,.
\eeq
The Jacobi equation now becomes
\beq 
\Xi^Z_\alpha\frac{\partial E^\nu_B}{\partial \omega^Z_\alpha}+ d_\beta \Xi^Z_\alpha \frac{\partial E^\nu_B}{\partial \omega^Z_{\alpha,\beta}}+ d_\gamma d_\beta \Xi^Z_\alpha \frac{\partial E^\nu_B}{\partial \omega^Z_{\alpha,\beta \gamma}}=0 \,.
\eeq 
Clearly now the terms
\beq 
\frac{\partial E^\nu_B}{\partial \omega^Z_\alpha}\text{ \ \ \ and\ \ \ \ }
\frac{\partial E^\nu_B}{\partial \omega^Z_{\alpha,\beta}}
\eeq 
do not vanish identically; indeed
we have
\begin{align*}
\Xi^Z_\alpha\frac{\partial E^\nu_B}{\partial \omega^Z_\alpha}= \Xi^Z_\alpha\left[\delta_{BA} c^A_{CZ}\omega^C_{\lambda, \mu}\eta^{\lambda \mu}\eta^{\alpha \nu}+\delta_{BA}c^A_{ZD}\omega^D_{\sigma,\mu}\eta^{\alpha \mu}\eta^{\sigma \nu}\right. +\\
\left.+\eta^{\alpha\mu}\eta^{\sigma\nu} c^D_{ZF}\omega^F_\sigma c^A_{BC}\omega^C_\mu\delta_{DA}+\eta^{\lambda\mu}\eta^{\alpha\nu}c^D_{EZ}\omega^E_\lambda c^A_{BC}\omega^C_\mu\delta_{DA}\right. +\\
\left.+\eta^{\lambda\alpha}\eta^{\sigma\nu}F^D_{\lambda\sigma}c^A_{BZ}\delta_{DA}\right] \,,
\end{align*}
and
\begin{align*}
d_\beta\Xi^Z_\alpha\frac{\partial E^\nu_B}{\partial \omega^Z_{\alpha,\beta}}=d_\beta\Xi^Z_\alpha\left[\delta_{BA}\eta^{\lambda \mu}\eta^{\sigma \nu}(c^A_{ZD}\delta^\alpha_\lambda\delta^\beta_\mu\omega^D_\sigma+c^A_{CZ}\delta^\alpha_\sigma\delta^\beta_\mu \omega^C_\lambda)\right. + \\
\left.+\eta^{\beta\mu}\eta^{\alpha\nu}\delta_{ZA}c^A_{BC}\omega^C_\mu-\eta^{\alpha\mu}\eta^{\beta\nu}\delta_{ZA}c^A_{BC}\omega^C_\mu\right] \,.
\end{align*}
The third term is analogous to the one in Maxwell case: 
\beq 
d_\gamma d_\beta \Xi^Z_\alpha \frac{\partial E^\nu_B}{\partial \omega^Z_{\alpha,\beta \gamma}}=\delta_{BZ}\eta^{\sigma \nu}\eta^{\kappa \tau}d_\kappa(d_\tau\Xi^Z_\sigma-d_\sigma\Xi^Z_\tau) \,.
\eeq 
Summing  up these terms and doing some straightforward calculations, we have, for any pair $(\nu,B)$,  
\begin{align*}
&  \eta^{\sigma \nu}\eta^{\alpha \beta}\left\{d_\beta\left[\left(d_\alpha\Xi^A_\sigma + c^A_{CZ}\Xi^Z_\sigma\omega^C_\alpha\right)\delta_{BA}\right]+\left[\left(d_\alpha\Xi^D_\sigma+\Xi^Z_\sigma c^D_{EZ}\omega^E_\alpha\right)\delta_{AD}\right]c^A_{BC}\omega^C_\beta\right. +\\
& - \left.d_\beta\left[\left(d_\sigma\Xi^A_\alpha + c^A_{CZ}\Xi^Z_\alpha\omega^C_\sigma\right)\delta_{BA}\right]-\left[\left(d_\sigma\Xi^D_\alpha+\Xi^Z_\alpha c^D_{EZ}\omega^E_\sigma\right)\delta_{AD}\right]c^A_{BC}\omega^C_\beta\right. + \\
& \left. + F^D_{\beta\sigma}c^A_{BZ}\Xi^Z_\alpha\delta_{AD}\right\} =0 \,.
\end{align*}

It is now noteworthy that we can further simplify the expression introducing a suitable induced connection. 

Let $(\phi^a)$ be a set of coordinates on the group $G$. Introducing right invariant vector fields $\rho_A$, we have $\rho_A=R^a_A(\phi)\partial_a$, where $\partial_a$ denotes $\frac{\partial}{\partial \phi^a}$, the standard local system of generators of vector fields on $G$. We will denote by $\overline{R}^A_a(\phi)$ the inverse matrix of $R^a_A(\phi)$; in an analogue way, using left invariant vector fields $\lambda_A$, we introduce the matrix $L^a_A(\phi)$ and its inverse $\overline{L}^A_a$. Moreover, we introduce $Ad^B_A(\phi)=\overline{R}^B_aL^a_A$, that is the {\em adjoint representation} of $G$ on $\mathfrak{g}.$ If we chose another system of fibered coordinates on $P$, $(x'^\nu,\phi'^b)$, we recall that
$ \omega'^{B}_\nu=\overline{J}^\mu_\nu\left(Ad^B_A(\phi)\omega^A_\mu-\overline{R}^B_a(\phi)\phi^a_\mu\right)$,
where $\overline{J}^\mu_\nu$ denotes the inverse of the Jacobian matrix of the change of coordinates in the base space. Then the components of a vertical vector field
satisfy the transformation rule
$ \Xi'^B_\nu=Ad^B_A(\phi)\Xi^A_\mu\overline{J}^\mu_\nu$.

Following a standard approach
we can see $\Xi$ as a section of a suitable bundle. Indeed, consider the fibered product $P\times_M L(M)$ where $L(M)$ is the frame bundle of $M$; $P\times_M L(M)$ is clearly a principal bundle with structure group $G\times GL(n)$, where $GL(n)$ is the general linear group of degree $n=dim(M)$. We introduce the vector space $V=\mathfrak{g}\otimes\mathbb{R}^n$ and the representation
\begin{align*}
\lambda:G\times GL(n) \times V &\to V\\
(\phi,J,\Xi^A_\nu)&\to \Xi'^B_\nu=Ad^B_A(\phi)\Xi^A_\mu\overline{J}^\mu_\nu  \,,
\end{align*}
by which we construct the bundle $B=(P\times_M L(M))\times_\lambda V$, which turns out to be associated  with $P\times_M L(M)$.  As well known, its sections are in one to one correspondence with vertical vector fields over $C_P$. 

Now we consider that a principal connection on $P\times_M L(M)$ is induced by any pair $(\omega,\Gamma)$, where $\omega$ is a principal connection on $P$ (for example, an extremal of the Yang-Mills Lagrangian) while $\Gamma$ is a principal connection on $L(M)$ (see \cite{Kol75,Kol79} and, for gauge-natural theories, \cite{FF03,FFP01}). In coordinates, if $\rho^\nu_\lambda$ are right invariant vector fields on $L(M)$,
\beq 
\Omega = dx^\mu \otimes\left(\partial_\mu-\omega^A_\mu\rho_A-\Gamma^\lambda_{\nu\mu}\rho^\nu_\lambda\right)
\eeq 
is a principal connection on $P\times_M L(M)$. However, since we are considering a manifold $M$ that admits a global Minkowskian metric, we get a connection on $L(M)$ with coefficients vanishing in a whole class of system of coordinates (the ones in which the metric is written as $\eta_{\mu \nu}$). 
We are already working in these coordinates, because we have required the metric to be expressed in diagonal form; then we can assume $\Gamma^\lambda_{\nu\mu}=0$. 

We thus induce a connection on any bundle associated with $P\times_M L(M)$; in particular, we have a connection on $B$ given by
\beq 
\tilde{\Omega}=dx^\mu\otimes\left(\partial_\mu-\omega^B_{\sigma\mu}(x,\Xi)\partial^{\sigma}_B\right) \,.
\eeq 
Now, taking into account that the coefficients of $\Gamma$ are assumed to vanish, 
\beq 
\omega^B_{\sigma\mu}(x,\Xi)=T^a_A\partial_a\lambda^B_\sigma(e,\Xi)\omega^A_\mu(x) \,,
\eeq 
where
\begin{itemize}
\item $e$ denotes the identity element of $G\times GL(n)$
\item $\lambda^B_\sigma$ denotes the ``components'' of the representation $\lambda$
\item $T_A=T^a_A\partial_a$ is a fixed basis of $\mathfrak{g}\cong T_{id}G$  ($id$ is the identity element of $G$).
\end{itemize}
Working out this expression in local coordinates we get
\beq \omega^B_{\sigma \mu}(x,\Xi)=-c^B_{AD}\Xi^D_\sigma\omega^A_\mu \,.
\eeq 
Therefore, by some careful  manipulations, we rewrite the Jacobi equation for the Yang-Mills Lagrangian on that specific background as
\bEq\label{JACOBI EQUATION}
\eta^{ \nu\sigma }\eta^{\beta \alpha }\left\{
\nabla_\beta\left[\left(\nabla_\alpha\Xi^A_\sigma-\nabla_\sigma\Xi^A_\alpha\right)\delta_{BA}\right]+
F^D_{\beta\sigma}c^A_{BZ}\Xi^Z_\alpha\delta_{AD}\right\}=0 \,,
\eEq
for any pair $(\nu,B)$.

We note that the above is comparable with the classical definition of a Jacobi operator \cite{AtBo83,Bou87}. It can be easily checked by writing down in our case the analogous  expression corresponding to $L_A=d_A^*d_A+ *[ *F, ]$ for the Jacobi operator given in \cite{AtBo83} page $553$ (we stress that the expression for the second variation in \cite{AtBo83,Bou87} is reproduced by our approach by taking the twice iterated variation by the same variation field, see also in particular \cite{GoSt73}).

The solution of the Jacobi equation defines the kernel $\mathfrak{k}$ of the Jacobi morphism $\cJ$, which, in particular, is characterized by Proposition \ref{H1} and Proposition \ref{H2}. 
Note that we have fixed an orthonormal basis for the Cartan-Killing metric; working with a specific group $G$ of course the equation can be further specialized writing down the structure constants. 

\subsection{Weak conservation laws associated with couples of Jacobi fields} \label{Jac_conservation}

As an example of application of the arguments discussed in this section, we calculate the current $\epsilon_{\tilde{\Xi}}\left(\Xi\rfloor E_n\left(\lambda_{YM}\right)\right)$ for two given Jacobi fields $\Xi$ and $\tilde{\Xi}$ along an extremal of the Yang-Mills Lagrangian $\lambda_{YM}$ on a Minkowskian background (see the previous section); details can be found in \cite{Ac17}. Being the vector fields vertical, the current has the form
\beq 
\epsilon_{\tilde{\Xi}}\left(\Xi\rfloor E_n\left(\lambda_{YM}\right)\right)=-J^{3}\tilde{\Xi}\rfloor p_1\mathcal{R}\left(d\left(\Xi\rfloor E_n\left(\lambda_{YM}\right)\right)\right)  \,.
\eeq 
Having denoted the components of the principal connection by $\omega^A_\mu$, in order to avoid confusion we will use in this example the notation $\theta^A_\mu, \theta^A_{\mu,\nu}, \theta^A_{\mu,\nu\rho}, \dots$ to indicate generators of contact forms. 
According to \cite{KrMu05}, we compute the residual operator  
 associated with the form
\begin{align*}
d\left(\Xi\rfloor E_n\left(\lambda_{YM}\right)\right)=\sum\limits^2_{|J|=0}\theta^A_{\mu,J}\wedge \eta^{\mu,J}_{A}&=\\
 = \left(\frac{\partial \Xi^B_\nu}{\partial \omega^Z_\rho}E^\nu_B+\Xi^B_\nu\frac{\partial E^\nu_B}{\partial \omega^Z_\rho}\right)\theta^Z_\rho\wedge ds+ &\left(\Xi^B_\nu\frac{\partial E^\nu_B}{\partial \omega^Z_{\rho,\xi}}\right)\theta^Z_{\rho,\xi}\wedge ds+\left(\Xi^B_\nu\frac{\partial E^\nu_B}{\partial \omega^Z_{\rho,\xi\tau}}\right)\theta^Z_{\rho,\xi\tau}\wedge ds \,,
\end{align*}
where we recall that $E^\nu_B$ denotes the coordinate expression of the Euler--Lagrange form.
We need to rewrite this form as $\sum\limits^2_{|I|=0}d_I(
\theta^A_\mu\wedge\zeta^{\mu,I}_A)$ with
\beq 
\zeta^{\mu,I}_A=\sum\limits^{2-|I|}_{|J|=0}(-1)^{|J|}{{|I|+|J|}\choose{|J|}}d_J\eta^{\mu,JI}_A \,.
\eeq 
Actually, we are interested only in $\zeta^{\mu,I}_A$ for $|I|=1$ or $|I|=2$; the case $|I|=0$ gives no contribution to the residual operator. We have, for $|I|=1$,
\beq 
\zeta^{\mu,I}_A=\eta^{\mu,I}_A-2d_\tau\eta^{\mu,\tau I}_A \,,
\eeq 
and, for $|I|=2$
\beq 
\zeta^{\mu,I}_A=\eta^{\mu,I}_A \,.
\eeq 
Then
\begin{align*}
\sum\limits^2_{|I|=1}d_I(\theta^A_\mu\wedge\zeta^{\mu,I}_A)=&d_\xi\left[\theta^Z_\rho\wedge\left(\Xi^B_\nu\frac{\partial E^\nu_B}{\partial \omega^Z_{\rho,\xi}}-2d_\tau\left(\Xi^B_\nu\frac{\partial E^\nu_B}{\partial \omega^Z_{\rho,\xi\tau}}\right)\right)ds\right] +\\
&+d_\tau d_\xi\left[\theta^Z_\rho\wedge\left(\Xi^B_\nu\frac{\partial E^\nu_B}{\partial \omega^Z_{\rho,\xi\tau}}\right)ds\right] \,,
\end{align*}
that can be rewritten as
\beq d_\xi\left\{\left[\Xi^B_\nu\frac{\partial E^\nu_b}{\partial \omega^Z_{\rho,\xi}}-d_\tau\left(\Xi^B_\nu\frac{\partial E^\nu_B}{\partial\omega^Z_{\rho,\xi\tau}}\right)\right]\theta^Z_\rho+\left(\Xi^B_\nu\frac{\partial E^\nu_B}{\partial \omega^Z_{\rho,\xi\tau}}\right)\theta^Z_{\rho,\tau}\right\}\wedge ds \,.
\eeq 
Consequently
\begin{align*}
\mathcal{R}\left(d\left(\Xi\rfloor E_n\left(\lambda_{YM}\right)\right)\right)=&-\left(\Xi^B_\nu\frac{\partial E^\nu_B}{\partial \omega^Z_{\rho,\xi}}-d_\tau\left(\Xi^B_\nu\frac{\partial E^\nu_B}{\partial \omega^Z_{\rho,\xi\tau}}\right)\right)\theta^Z_\rho\wedge ds_\xi + \\
& - \left(\Xi^B_\nu\frac{\partial E^\nu_B}{\partial \omega^Z_{\rho,\xi\tau}}\right)\theta^Z_{\rho,\tau}\wedge ds_\xi \,,
\end{align*}
and the current is
\begin{align*}
\epsilon_{\tilde{\Xi}}\left(\Xi\rfloor E_n\left(\lambda_{YM}\right)\right)=&\left[\Xi^B_\nu\frac{\partial E^\nu_B}{\partial \omega^Z_{\rho,\xi}}\tilde{\Xi}^Z_\rho-d_\tau\left(\Xi^B_\nu\frac{\partial E^\nu_B}{\partial \omega^Z_{\rho,\xi\tau}}\right)\tilde{\Xi}^Z_\rho+\Xi^B_\nu\frac{\partial E^\nu_B}{\partial \omega^Z_{\rho,\xi\tau}}d_\tau\tilde{\Xi}^Z_\rho\right] ds_\xi \,,
\end{align*}
where
\begin{align*}
\frac{\partial E^\nu_B}{\partial \omega^Z_{\rho,\xi}}=& \left[\delta_{BA}c^A_{ZD}\omega^D_\sigma\left(\eta^{\rho\xi}\eta^{\sigma\nu}-\eta^{\sigma\xi}\eta^{\rho\nu}\right)+\delta_{ZA}c^A_{BC}\omega^C_\sigma\left(\eta^{\rho\nu}\eta^{\sigma\xi}-\eta^{\rho\sigma}\eta^{\xi\nu}\right)\right]  \,,\\
\frac{\partial E^\nu_B}{\partial \omega^Z_{\rho,\xi\tau}}=& \delta_{BZ}\left(\eta^{\xi\tau}\eta^{\rho\nu}-\eta^{\rho(\tau}\eta^{\xi)\nu}\right) \,.
\end{align*}
(the brackets on the superscripts denote symmetrization).
Substituting we can write, for the coefficients of the current,
\begin{align*}
& \left(\Xi^B_\nu d_\tau\tilde{\Xi}^Z_\rho-d_\sigma\Xi^B_\nu\tilde{\Xi}^Z_\rho\right)\left(\eta^{\xi\tau}\eta^{\rho\nu}-\eta^{\rho(\tau}\eta^{\xi)\nu}\right)\delta_{BZ} + \\
&+\Xi^B_\nu\tilde{\Xi}^Z_\rho\delta_{BA}c^A_{ZD}\omega^D_\sigma\left(\eta^{\sigma\nu}\eta^{\rho\xi}-\eta^{\sigma\xi}\eta^{\rho\nu}\right)+\Xi^B_\nu\tilde{\Xi}^Z_\rho\delta_{ZA}c^A_{BC}\omega^C_\sigma\left(\eta^{\rho\nu}\eta^{\sigma\xi}-\eta^{\sigma\rho}\eta^{\xi\nu}\right) \,.
\end{align*}
Denoting with square brackets the anti-symmetrization, we can formulate this as
\begin{align*}
\eta^{\rho[\xi}\eta^{\sigma]\nu}\delta_{BA}c^A_{ZD}\omega^D_\sigma\left(\Xi^B_\nu\tilde{\Xi}^Z_\rho+\Xi^Z_\nu\tilde\Xi^B_\rho\right)+\left(\eta^{\xi\sigma}\eta^{\rho\nu}-\eta^{\rho(\sigma}\eta^{\xi)\nu}\right)\left(\Xi^B_\nu d_\sigma\tilde{\Xi}^Z_\rho\delta_{BZ}\right. + \\
\left. - d_\sigma\Xi^B_\nu \tilde{\Xi}^Z_\rho\delta_{BZ}+\Xi^B_\nu\tilde{\Xi}^Z_\rho\delta_{ZA}c^A_{BD}\omega^D_\sigma-\Xi^B_\nu\tilde{\Xi}^Z_\rho\delta_{BA}c^A_{ZD}\omega^D_\sigma\right) \,.
\end{align*}
In conclusion, the current is given by
\begin{equation}
\begin{aligned}
\epsilon_{\tilde{\Xi}}\left(\Xi\right.&\left.\rfloor E_n\left(\lambda_{YM}\right)\right)=\left[\eta^{\rho[\xi}\eta^{\sigma]\nu}\delta_{BA}c^A_{ZD}\omega^D_\sigma\left(\Xi^B_\nu\tilde{\Xi}^Z_\rho-\Xi^Z_\rho\tilde\Xi^B_\nu\right)+\right. \\
&\left.\left(\eta^{\xi\sigma}\eta^{\rho\nu}-\eta^{\rho(\sigma}\eta^{\xi)\nu}\right)\left(\Xi^B_\nu\nabla_\sigma\left(\tilde\Xi^Z_\rho\delta_{ZB}\right)-\tilde{\Xi}^Z_\rho\nabla_\sigma\left(\Xi^B_\nu\delta_{BZ}\right)\right)\right]ds_\xi \,.
\end{aligned}
\end{equation}
An interpretation of this current in terms of invariance properties of the set of extremals and related  Noether currents has been obtained in \cite{AcPa19}.

\section*{Acknowledgements}
Research partially supported by Department of Mathematics - University of Torino through the  projects PALM$\_$RILO$\_16\_ 01$ and FERM$\_$RILO$\_17\_ 01$ (MP).
 and written under the auspices of GNSAGA-INdAM. 
 The first author (LA) is also supported by a NWO-UGC project, Grant BM$.00193.1$.


\end{document}